\documentclass[man]{apa6}
\usepackage[]{graphicx}
\usepackage[]{color}
\makeatletter
\def\maxwidth{ %
  \ifdim\Gin@nat@width>\linewidth
    \linewidth
  \else
    \Gin@nat@width
  \fi
}
\makeatother

\definecolor{fgcolor}{rgb}{0.345, 0.345, 0.345}

\usepackage{framed}
\makeatletter
 {\par\unskip\endMakeFramed%
 \at@end@of@kframe}
\makeatother

\definecolor{shadecolor}{rgb}{.97, .97, .97}
\definecolor{messagecolor}{rgb}{0, 0, 0}
\definecolor{warningcolor}{rgb}{1, 0, 1}
\definecolor{errorcolor}{rgb}{1, 0, 0}
\newenvironment{knitrout}{}{} % an empty environment to be redefined in TeX

\usepackage{alltt} % man for manuscript format, jou for journal format, doc for standard LaTeX document format
\usepackage[natbibapa]{apacite} % Divine intervention help you if you need to use a different citation package.
\usepackage[american]{babel}
\usepackage[utf8]{inputenc}
\usepackage{csquotes}

\usepackage{url}   % this allows us to cite URLs in the text
\usepackage{graphicx}   % allows for graphic to float when doing jou or doc style
\usepackage{verbatim}   % allows us to use \begin{comment} environment
\usepackage{caption}
\usepackage{pdflscape}

\usepackage{epstopdf}
\usepackage{amsmath}
\usepackage{float}
\usepackage{array,multirow,makecell}
\usepackage{lscape}
\usepackage{tabularx}

\usepackage{verbatim}
\makeatletter
\def\verbatim@font{\linespread{1}\normalfont\ttfamily}
\makeatother

\usepackage{lineno, blindtext}

\makeatletter
\def\verbatim@font{\linespread{1}\normalfont\ttfamily}
\makeatother

\title{A Tutorial for Joint Modeling of Longitudinal and Time-to-Event Data in \texttt{R}}
\shorttitle{A Tutorial for Joint Modeling in \texttt{R}}

\fiveauthors{Sezen Cekic}{Stephen Aichele}{Andreas M. Brandmaier}{Ylva K\"{o}hncke}{Paolo Ghisletta}

\fiveaffiliations{University of Geneva, Geneva, Switzerland}{University of Geneva, Geneva, Switzerland}{Max Planck Institute for Human Development, Berlin, Germany\\Max Planck UCL Centre for Computational Psychiatry and Ageing Research, Berlin, Germany}{Max Planck Institute for Human Development, Berlin, Germany}{University of Geneva, Geneva, Switzerland\\Swiss Distance Learning University, Switzerland\\ Swiss National Centre of Competence in Research LIVES\\Universities of Lausanne and of Geneva, Switzerland}

\leftheader{Cekic et al.}

\authornote{Correspondence concerning this article should be addressed to Sezen Cekic, Faculty of Psychology and Educational Sciences, University of Geneva, Boulevard du Pont d'Arve 40, CH-1211 Geneva 4, Switzerland. E-mail: sezen.cekic@unige.ch}
\note{\today}

\journal{} %\journal{Fictional Journal of Awesome} % for jou format
\volume{} %\volume{5,1,5-10}% Volume, number, pages; typeset in the top left header in jou and doc modes, underneath the content of %\journal


\abstract{In biostatistics and medical research, longitudinal data are often composed of repeated assessments of a variable (e.g., blood pressure or other biomarkers) and dichotomous indicators to mark an event of interest (e.g., recovery from disease, or death). Consequently, joint modeling of longitudinal and time-to-event data has generated much interest in these disciplines over the previous decade. In psychology, too, often we are interested in relating individual trajectories (e.g., cognitive performance or well-being across many years) and discrete events (e.g., death, diagnosis of dementia, or of depression). Yet, joint modeling are rarely applied in psychology and social sciences more generally. This tutorial presents an overview and general framework for joint modeling of longitudinal and time-to-event data, and fully illustrates its application in the context of a behavioral (cognitive aging) study. We discuss practical topics, such as model selection and comparison for both longitudinal and time-to-event data, choice of joint modeling parameterization, and interpretation of model parameters. To do so, we examined seven frequently used packages for joint modeling in the \texttt{R} language and environment. We concluded that of these, \textbf{JMbayes} is especially attractive due to its flexibility, its various parameterizations of the association structure, and for its powerful and fully Bayesian implementation. We make available the \texttt{R} syntax to apply the \textbf{JMbayes} package within our example.}

\ccoppy{} %\ccoppy{Copyright notice, etc.; typeset in the top right header of page 1 (jou and doc modes only)}
\IfFileExists{upquote.sty}{\usepackage{upquote}}{}
\begin{document}

\maketitle

\section{Introduction}
\label{Introduction}

In many research settings, it is very common to record longitudinal observations that correspond to responses on continuous measures alongside dichotomous indicators to mark the occurrence of an event of interest. A prototypical example in clinical research is the repeated assessment of biological measures (e.g., blood pressure, antibody affinity, cholesterol level) that may relate to an event such as death, recovery from a disease, or disease diagnosis. In psychological research, much interest evolves around the association between long-term individual trajectories of cognitive performance and imminent death \citep[e.g., terminal decline or terminal drop hypotheses; ][]{kleemeier1962,riegel1972}.

Methods for the separate analysis of such outcomes are well established in the literature, and these frequently include the use of mixed-effects models for the longitudinal portion of the data and the Cox proportional hazards model for the time-to-event data \citep{Laird_1982,cox1972}. However, the repeated measurements of the same individual are most likely related to the event, making them \textit{endogenous}. This means that the measurement value at a given time point is informative about the future occurrence (or non-occurrence) of the event of interest. Likewise, the longitudinal trajectory is related to, and shaped by, the occurrence of the event. Separate analyses of such data cannot account for this endogeneity and cannot inform about the associative strength between the two. A tentative solution to this problem is a two-step approach, whereby estimated parameters from the mixed-effects models are used as covariates in the time-to-event model. Yet, this approach has been shown to suffer from increased bias in estimation and loss of efficiency \citep{prentice1982covariate,rizopoulos_book}. Joint models \textit{simultaneously} estimate both longitudinal and time-to-event components and are better suited for analyzing such data because they estimate \textit{jointly} the relative risk of the time-to-event outcome contingent upon the longitudinal outcome \citep{Wulfsohn1997,Song_2002,Henderson2000,hogan1997mixture,tsiatis2004joint,gould_joint_2015,
Papageorgiou_2019}.

Joint modeling is increasingly used in medical research \citep[see for example][]{Rizopoulos_EuroIntervention,nunez2014red,thabut2013survival,abdi2013impact,mauritz2011usefulness,sene_shared_2014,brown2003bayesian,he2016joint}. However, joint models remain conspicuously absent in psychological and behavioral research, despite the  proliferation of data that could be analyzed more effectively using this powerful methodology, rather than by using a two-step estimation procedure. For example, given world-wide increases in human life expectancy (and corresponding increases in populations of elders), it will become increasingly important to understand how long-term changes in cognitive, personality, emotional, and other individual behavioral characteristics are related both to specific health-related outcomes (depression, stroke, dementia) and to longevity, or risk of death \citep{world2015world}. We found only a few behavioral studies using joint models to evaluate such associations \citep{Ghisletta_2006,ghisletta2008application,McArdle_2005,muniz2011,Muniz_2018} and we believe that the lack of an accessible and clear guidance for implementing such models remains an obstacle preventing non-expert users from applying them skillfully, or from adopting them at all.

This tutorial is therefore aimed at elucidating features of the joint modeling framework for longitudinal and time-to-event data for users familiar with mixed-effects and time-to-event (a.k.a. survival) models. We initially reviewed the seven currently available joint-modeling packages in the open-source \texttt{R} environment for statistical computing \citep{R}. The seven packages are \textbf{JM}, \textbf{JMbayes}, \textbf{joineR}, \textbf{lcmm}, \textbf{frailtypack}, \textbf{rstanarm} and \textbf{bamlss}, which are described in tables \ref{Table JM}, \ref{Table JMbayes}, \ref{Table joineR} , \ref{Table lcmm}, \ref{Table frailtypack}, \ref{Table rstanarm} and \ref{Table bamlss}, respectively. We concluded that, at present, the \textbf{JMbayes} package \citep{JMbayes} is the most comprehensive and extensible, and thus chose it as the basis for this tutorial. Note that a range of tools allowing one to perform joint modeling are available for other standard statistical software such as SAS/STAT, Stata, WinBUGS, and JAGS \citep[see][and references therein for more details]{Papageorgiou_2019}.

\textbf{JMbayes} provides a framework wherein one or more characteristics from the longitudinal submodel can be flexibly parameterized as linear predictors in the time-to-event submodel. This flexibility is important because the relation between the continuous variable and the event of interest may take several functional forms, each of which may imply a different theoretical interpretation. For example, the longitudinal process can be specified in relatively simple terms (e.g., individual random effects about the intercept and the linear slope within a mixed-effects submodel), or in more complex non-linear spline effects, each of which can be linked to the time-to-event process with any of several association structures. \textbf{JMbayes} also allows for inclusion of multivariate longitudinal processes and for stratification criteria in the time-to-event submodel. Also, estimation of joint models, which are inherently complex, can be computationally intensive. Per its name, the \textbf{JMbayes} package relies on a Bayesian analytical approach, which is very attractive for applications where classical maximum likelihood estimation
%may depend on overly strict modeling assumptions and/or simply fail to converge. Furthermore, maximum likelihood estimation
often encounters nonidentifiability issues yielding unreliable inferential conclusions. Indeed, Bayesian analysis using Markov chain Monte Carlo (MCMC) estimation can yield very precise point estimates without reliance on asymptotic assumptions \citep{robert2004monte}. \textbf{JMbayes} also provides useful diagnostic measures for evaluating MCMC convergence and for evaluating relative model fit.

We note that the aim of this tutorial is not to replace the \textbf{JMbayes} documentation, which provides in-depth methodological detail and supplementary \texttt{R} code \citep{JMbayes,rizopoulos2017package}. Here, our goal is to complement this information with a more user-oriented, applied perspective, so that non-expert users may successfully carry out analyses using joint models. For the sake of brevity and clarity, we do not provide an exhaustive treatment of all features of joint modeling from the applications, but we nevertheless briefly discuss some of them at the end of the tutorial. We begin by describing the statistical framework for joint modeling. We then discuss required software packages, describe the data set used for this tutorial, and finally present example analyses to illustrate the use of joint modeling for behavioral data analysis. We provide the data and the full syntax of our application as Supplementary material.

%The tutorial is organized as follows: We first explained the theoretical framework for the joint model; then, we presented the JMbayes R package; Then we described the dataset used for illustrating practical applications of the JMbayes package; Section \ref{examples} presents these practical applications of the joint modeling framework; Section \ref{discussion} points to additional features that we did not touch upon due to space constraints.

\section{Statistical Framework}
\label{Statframe}

The joint model, first described by \cite{Wulfsohn1997}, consists of two submodels: a mixed-effects submodel for the longitudinal data and a time-to-event submodel for the time-to-event data. The two submodels are joined by allowing the time-to-event submodel to depend on \textit{some characteristics} of the longitudinal submodel potentially in various ways. The defining feature of the joint model is that longitudinal and time-to-event data are modeled \textit{simultaneously} with respect to a \textit{conditional joint density}, rather than separately with two marginal and independent densities \citep[see e.g.,][for further technical details]{Rizopoulos_JASA}. Consequently, the effects of the longitudinal submodel on the time-to-event submodel are explicitly specified, and, at the same time, the parameter estimates of the former are also conditioned on the latter submodel (i.e., the longitudinal submodel is more accurately estimated by including the time-to-event data than without them, if endogeneity occurs). We discuss first the longitudinal submodel, then the time-to-event submodel, and finally several parametrization options to associate the two within the joint model.

\subsection{The Mixed-Effects Submodel}
\label{mixed_model}

Longitudinal data here consist of repeated measurements of the same outcome for each study participant over a given period of time. Repeated measurements for the same individual are typically correlated, and therefore each individual in the population is expected to have his or her own subject-specific response pattern over time. The mixed-effects model \citep{Laird_1982} accounts for this between person variability by estimating person specific random effects around the parameters that are invariant across individuals (fixed effects).

Let $y_{i}(t)$ denote \textit{the measurement} on $y$ for individual $i$ ($i=1, \dots, N$) at time $t$ ($t=1, \dots, T^{*}_i$), where $T^{*}_i$ is the time-to-event for individual $i$. The model assumes that $y_{i}(t)$ is the \textit{observed} measured value, which deviates from the \textit{true unobserved value} $\mu_i(t)$ by the amount of error $\epsilon_i(t)$, where $\epsilon_i(t) \sim \mathcal{N}(0, \sigma^{2})$, normally distributed around 0 with variance $\sigma^2$. The longitudinal submodel can then be written as
\begin{equation}
\begin{aligned}
\mu_i(t) &= X_i(t) \beta + Z_i(t) b_i, \quad \text{or}\\
 		& = f(X_i(t))+f_i(Z_i(t)),
\end{aligned}
\label{Longitudinal_model}
\end{equation}
where $b_i$ denotes the vector of random effects for individual $i$ with $b_i \sim \mathcal{N}(0, D)$, and $D$ is the variance-covariance matrix of the random effects (typically a $2 \times 2$ unstructured matrix with random intercept and random slope effects). $X_i(t)$ and $Z_i(t)$ represent the design matrix containing the predictors for the fixed-effects regression coefficients $\beta$ and for the random-effects regression coefficients $b_i$, respectively (typically $b_{0i}$ and $b_{1i}$ for random intercept and slope effects).

The second line of Equation \eqref{Longitudinal_model} represents the notation when $\mu_i(t)$ is modeled non-parametrically via a spline approach (illustrated below). $f(\cdot)$ and $f_i(\cdot)$ represent the spline function for the fixed and, respectively, random components.

\subsection{The Time-to-Event Submodel}
\label{survival model}

Time-to-event, or survival, analysis refers to statistical methods for analyzing time-to-event data. An event time is the time elapsed up to the occurrence of an event of interest (such as death or disease diagnostic), given that it has not previously occurred yet. Time-to-event data are characterized by the fact that the event of interest may not have occurred for every participant during the duration of the study. This particular type of missing data is analogously treated as data of a participant who drops out before the end of the study, and both scenarios constitute \textit{right censorship}. Time-to-event analysis is particularly appropriate to analyze right-censored data.

There are different models for time-to-event data \citep{kaplan1958nonparametric}. Probably, the most popular is the semi-parametric Cox proportional hazard (PH) model \citep{cox1972}, which is a regression model wherein covariates are assumed to have a multiplicative effect on the hazard for an event. The Cox PH model has the particularity of estimating the baseline hazard function non-parametrically. If there is a theoretical reason to suppose that the baseline hazard function has a particular parametric form (typically Weibull, Gamma or Exponential), a parametric model can be specified. Most often, however, an analyst ignores the exact form of a hazard function. The Cox PH model has been shown to approximate well unknown parametric hazard functions \citep{fox2002cox}.

The Cox PH submodel can be formulated as
\begin{equation}
h_i(t) = h_0 (t) \exp (\gamma^T w_i),
\label{Survival model1}
\end{equation}
where $h_0 (t)$ is the baseline hazard function at time $t$ and $w_i$ is a vector of baseline predictors with corresponding regression coefficients $\gamma$. The submodel has two distinct parts. First, $h_0 (t)$ describes the hazard (e.g., risk of the event occurring, given it has not yet occurred) when all covariates $w_i$ have value zero. This baseline hazard function is assumed invariant across all individuals ($h_0 (t)$ does not depend on $i$). Second, the effect parameters $\gamma^T w_i$ describe how the hazard varies as a function of the explanatory covariates $w_i$.

\subsection{The Joint Model}
\label{Joint Models}

Joint models formally associate the longitudinal and time-to-event processes through \textit{shared} parameters  \citep{Wulfsohn1997,Henderson2000,rizopoulos_book,ibrahim2010basic}.
This framework therefore models the hazard of experiencing the focal event as dependent on a subject-specific characteristic of its longitudinal trajectory. More specifically, we have
\begin{equation}
\begin{split}
\begin{aligned}
h_i(t) &= \lim_{\Delta t\to 0} \frac{\texttt{Pr}\{t \leq T^{*}_i + \Delta t | T^{*}_i \geq t, \mathcal{M}_i(t),w_i\}}{\Delta t}\\
 &= h_0 (t) \exp \big(\gamma^T w_i +f\{\mu_i(t),b_i,\alpha\}\big),
 \end{aligned}
\end{split}
\label{Survival_model_general}
\end{equation}
where $\mathcal{M}_i(t) = \{ \mu_i(s), 0 \leq s < t \}$ denotes the history of the true unobserved
longitudinal process $\mu_i(s)$ up to time point $t$, where $s$ is a time point prior to $t$.

$w_i$ is a vector of (possibly time-varying) covariates  with corresponding regression coefficients $\gamma$, and $b_i$ is the vector of random effects for individual $i$ defined in Equation \eqref{Longitudinal_model}. Parameter vector $\alpha$ quantifies the association between a priori \textit{selected features} of the longitudinal process and the hazard for the event at time $t$. Various options for the function $f(\cdot)$ are possible and lead to different forms of the time-to-event submodel. We discuss now the three most frequently used association functions in the joint modeling literature \citep[see for example][]{Wulfsohn1997,Henderson2000,GuoCarlin,tsiatis2004joint,Rizopoulos_JASA,sene_shared_2014,
gould_joint_2015,he2016joint,Papageorgiou_2019}.

\subsubsection{The ``current value'' association}
\label{CV}
The first association structure between the longitudinal and the time-to-event submodels we discuss is known as the ``current value'', and assumes that for individual $i$, the true value $\mu_i(t)$ of the longitudinal measure at time $t$ is predictive of the risk $h_i(t)$ of experiencing the event at that same time $t$. The Cox PH submodel with this association structure is written as
\begin{equation}
\begin{split}
h_i(t) = h_0 (t) \exp \big(\gamma^T w_i+ \alpha_1 \mu_i(t)\big).
\end{split}
\label{Survival model_p1}
\end{equation}
This equation associates $\mu_i(t)$, as defined in Equation \eqref{Longitudinal_model}, with $h_i(t)$ at each time point $t$
. The reliable estimate of $\mu_i(t)$ is of chief importance, and is obtained by correctly specifying the design matrices $X_i(t)$ and $Z_i(t)$ in Equation \eqref{Longitudinal_model}. This means that we need to specify a longitudinal submodel that accurately describes the repeatedly observed measurements $y_i(t)$.

The coefficient $\alpha_1$ provides a measure of the strength of the association between $\mu_i$ at time $t$ and the time-to-event process at the same time. More precisely, each one-unit increase on the current value of $\mu_i(t)$ is associated with an $\exp (\alpha_1)$-fold increase in a participant's risk of experiencing the event at that same time $t$, given the event has not occurred prior to $t$. Note that this parameterization supposes that the associative parameter $\alpha_1$ between the longitudinal value and the event risk is the same across all individuals. However, the overall hazard $h_i(t)$ may vary across individuals as a function of the other subject-specific covariates $w_i$.

\subsubsection{The ``current value plus slope '' association}
\label{Current value plus slope}

The ``current value'' parameterization, however, does not distinguish between individuals who have, at a specific time point, an equal longitudinal score (an equal $\mu_i(t)$), but who differ in their \textit{rate of change} of this score (e.g., one subject may have an increasing trajectory, whereas another a decreasing trajectory). The second association structure between the time-to-event and the longitudinal submodels is known as the ``current value plus slope'' parameterization. This extends the previous structure by adding the rate of change of the measurement at time $t$, estimated as the derivative of $\mu_i(t)$ w.r.t. the time metric, as follows:
\begin{equation}
\mu_i^{'}(t)=\frac{d\mu_i(t)}{dt},
\label{derivative}
\end{equation}
where $\mu_i^{'}(t)$ is the derivative of $\mu_i(t)$ with respect to the time variable $t$. Practically, the quantity $\mu_i^{'}(t)$ has to be calculated ``by hand'' by the user. The reader is referred to any basic mathematical literature for calculus derivative rules. For worked examples, see Equation \eqref{derivative_MLSC}.

The corresponding submodel of this association structure has the form
\begin{equation}
\begin{split}
h_i(t) = h_0 (t) \exp \big(\gamma^T w_i+ \alpha_1 \mu_i(t)+\alpha_2 \mu_i^{'}(t)\big),
\end{split}
\label{Survival model_p2}
\end{equation}
where the hazard of experiencing the event at time $t$ is now assumed to be associated with both the value of $\mu_i(t)$ and its rate of change (i.e., slope) $\mu_i^{'}(t)$, at time $t$ (see Equation \eqref{derivative}).

The coefficients $\alpha_1$ and $\alpha_2$ express the strength of the association between the value and rate of change of the (true) subject trajectory at time $t$ and the time-to-event process at the same time. For individuals having the same level of $\mu_i(t)$, the hazard ratio for a one-unit increase of the current rate of change $\mu_i^{'}(t)$ (i.e., the trajectory) is $\exp (\alpha_2)$. Identically, for individuals having the same rate of change $\mu_i^{'}(t)$, the hazard ratio for a one-unit increase of the current value $\mu_i(t)$ is $\exp (\alpha_1)$.

\subsubsection{The ``shared random-effects'' association}
\label{SRE}
The third association structure between the time-to-event and the longitudinal submodels is a ``shared random-effects'' parameterization \citep{Wulfsohn1997}. This parameterization includes only the random effects from the longitudinal submodel in Equation \eqref{Longitudinal_model} as linear predictors of the relative risk submodel. For a simple random-effects structure (i.e., random intercept and slope effects), the random effects represent the individual deviations from the sample average intercept and slope values, and the association parameters therefore reflect the change in the log hazard for a one-unit change in these deviations. For more elaborate structures, such as nonlinear parametrization, associative parameters should be interpreted with caution \citep{rizopoulos_book,Rizopoulos_JASA,gould_joint_2015}.
The ``shared random-effects'' parameterization postulates that, compared to the general tendency, individuals who have a lower (or higher) measure at baseline (i.e., random intercept) or who show a steeper decrease (or increase) in their longitudinal trajectories (i.e., random slope) are more or less likely to experience the event.
The ``shared random-effects'' submodel can be written as
\begin{equation}
h_i(t) = h_0 (t) \exp \big(\gamma^T w_i +\alpha^{T}b_i\big).
\label{Survival model_p3}
\end{equation}
This parameterization is computationally simpler than the ``current value'' and the ``current value plus slope'' parameterizations, because the associative part $\alpha^{T}b_i$ in Equation \eqref{Survival model_p3} is time-independent. Indeed, when a simple linear random-intercept and/or random-slope structure is assumed for the longitudinal submodel in \eqref{Longitudinal_model}, individual specificities are characterized by random effects $b_i$ witch do not depend on time $t$. This means that for a specific individual $i$, random effects $b_i$ are unique, regardless of the time $t$, contrary to the current value $\mu_i(t)$ and the current slope $\mu_i(t)^{'}$ \citep[see e.g.,][for further details]{Rizopoulos_JASA}.

\section{Software}
\label{software}

The \textbf{JMbayes} package in \texttt{R} has a basic model fitting function called \texttt{jointModelBayes($\cdot$)}, which accepts as its main arguments a mixed-effects submodel object fitted by the function \texttt{lme($\cdot$)} from package \textbf{nlme} \citep{nlme} and a time-to-event submodel object fitted by a Cox PH model with the function
\texttt{coxph($\cdot$)} from the package \textbf{survival} \citep{survival-package}. Thus, both these packages need to be installed to use \textbf{JMbayes}. \textbf{JMbayes} also requires that JAGS \citep[Just Another Gibbs Sampler;][]{plummer2004jags} be installed in order to perform MCMC estimation. We show the \texttt{R} syntax to properly install the \textbf{JMbayes} package in Section Code 1 lines 1-4 in the Computer Code Section.

Both the \textbf{nlme} and \textbf{survival} packages are commonly used for fitting linear mixed-effects and, respectively, Cox PH submodels within the \texttt{R} environment. We hence do not discuss them at length here. We refer the interested readers to the original articles \citep{nlme,survival-package}.

\section{Empirical Example}
\label{The Dataset}

We illustrate the joint modeling approach by applying \textbf{JMbayes} to data from a longitudinal study of cognitive change in adults \citep{Rabbitt2004}.

\subsection{The MLSC Dataset}
The data come from the Manchester Longitudinal Study of Cognition (MLSC) \citep{Rabbitt2004}. The MLSC study is a $20$-plus-year study consisting in over $6200$ adult individuals, with initial ages from $42$ to $97$. Five domains of cognitive ability were examined, but here, for simplicity, we focus only on processing speed (PS), defined as the speed at which we can compute simple but abstract tasks. This ability has been shown to decrease markedly during the lifespan, to induce general decline in more elaborate cognitive abilities (such as memory or verbal capacities), and to relate to physical health in older adults \citep{salthouse1996}. Here, we investigate whether an individual's level of and change in PS performance in adulthood predicts their risk of death.

PS was assessed up to four times at approximately 3-year intervals. Relations between cognition and survival have been the focus of several previous MLSC analyses \citep[see][and references therein for further details]{aichele2015life}. For the current illustration, we use data from female participants aged between $50$ and $90$ years only. This gave an effective N = $2780$ participants, of whom $1858$ died during the study, who were tested on average twice. Other MLSC variables included in the current analyses were chronological age at each of the longitudinal assessments and at study entry for the longitudinal submodel, and smoking status and chronological age at study entry, time in the study, and survival status for the time-to event submodel.

% DONE
% and??
% age at study entry: j'ai enlevé factor mais savoir que s'en est un...

\begin{figure}[!tb]
\centering
%\hspace*{-1in}
\includegraphics[height=9cm,keepaspectratio]{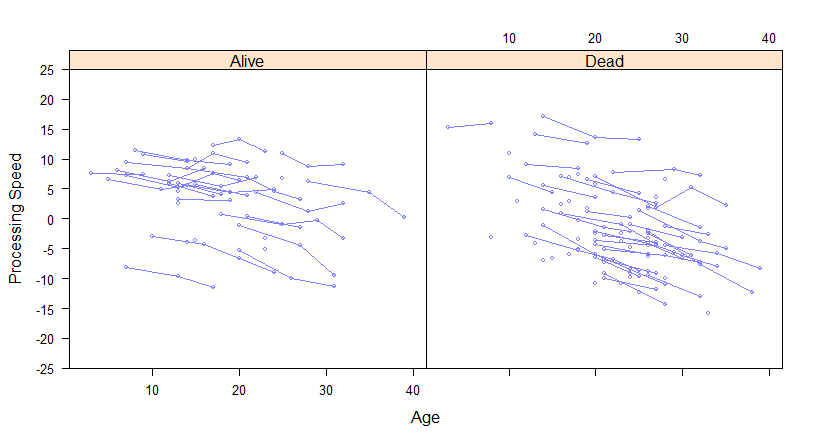}
\caption{Subject-specific longitudinal trajectories of PS cognitive performance for $100$ randomly selected women in the MLSC database with and without an event.}
\label{mlsc_AD}
\end{figure}

\subsection{MLSC Analysis Approach}

We hypothesize that mortality may explicitly depend on previous PS trajectories, whereby individuals with lower scores and/or steeper trajectories should have a greater hazard of dying than those performing better. We also posit that, given the PS-mortality association, adequate description of cognitive change in adult individuals should statistically take into account their (future) mortality information. A previous survival analysis of these data showed that the proportional hazards assumption was not met when comparing mortality risk for male and female participants \citep{aichele2015life}. Therefore, sex-specific analyses are warranted. For simplicity, we limit the current analyses to data from female MLSC participants. Before estimating the joint model, we perform initial model selection steps independently for the longitudinal and for the time-to-event submodels.

We first select a submodel for the longitudinal part of the data. Based on a previous literature search on the use of the mixed-effects model in cognitive aging research \citep{ghisletta2019}, we tested six submodels: the first with fixed and random-effects for the intercept and the linear slope (to describe change over time), the second adding to the first a quadratic slope (change trajectories with a single curvature), the third adding to the second a cubic slope (change trajectories with two curvatures), and the fourth to the sixth with spline effects of time and with different numbers of nodes. Of the six submodels, based on statistical criteria, we choose the one that provides the most satisfying description of the PS trajectories.

We then, separately, select the optimal time-to-event submodel. To do so we rely on the Cox PH model that relates smoking status and initial age to the timing and occurrence of death. For each covariate we test the PH assumption and then retain the submodel that best describes the mortality process based on the selected covariates. Finally, we combine both submodels in joint modeling, and compare four association structures between the two submodels.

\subsubsection{Longitudinal model selection}
\label{long_mlsc}

As required by the \textbf{JMbayes} package, we fit the longitudinal submodel with the \texttt{lme($\cdot$)} function from the \textbf{nlme} package \citep{nlme}. This longitudinal submodel must be specified in terms of its fixed and random components. For both components, linear and/or higher-degree polynomial terms are allowed (e.g., linear, quadratic or cubic change). More complex parameterizations of \textit{nonlinear} effects are also feasible, such as achieved through the flexible \textit{spline} methodology \citep[see the second line in Equation \eqref{Longitudinal_model} and cf.][]{rizopoulos_book}.

%Although the spline methodology was not applicable in our example due to problems with model convergence, most probably due to the low maximum number of repeated assessments (i.e., 4),
The spline approach has gain much popularity in recent years because of its flexibility in describing subject-specific longitudinal trajectories that follow highly nonlinear functions and that, individually, deviate from the average sample function. A spline is a piecewise function composed of polynomials adjusted to data within adjacent intervals, separated by equidistant points called nodes. The degree of the spline is defined by the polynomial of the highest degree used. For instance, if the function fitted to the data within an interval is a straight lines, the spline is said to be of degree $1$. Cubic splines are a popular specification for splines that entails polynomials of degree $3$ \citep{hastie1987generalized}. Within the \texttt{lme($\cdot$)} function, nonlinear effects can be specified with the subfunction \texttt{ns($\cdot$)} from the \textbf{splines} package. The \texttt{ns}($\cdot$) function uses natural cubic splines that estimate a smooth functions of the predictors \citep[functions $f(\cdot)$ and $f_i(\cdot)$ in the second line of Equation \eqref{Longitudinal_model};][]{hastie_statistical_2017}. With natural cubic splines, the data within each interval are fitted by a cubic spline, and splines from neighboring intervals must respect two conditions: at each node, the first and the second derivatives of the splines are continuous, meaning that, for any two adjacent intervals, the finishing portions of the left spline smoothly joins the starting portion of the right spline, thereby giving the visual impression of a single, continuos line passing through both intervals.

The user must select the optimal number of nodes, which is a rather arbitrary component of splines. However, this value should be less than or equal to the number of repeated measures. To this end, there is no automatic selection procedure implemented in the \textbf{JMbayes} package \citep[in contrast to the \textbf{mgcv} package, for example;][]{wood2006}. We suggest to first investigate the number of nodes for the fixed part of the longitudinal submodel, before proceeding to explore that of random part \citep[when both components are modeled using splines, as proposed in][]{JMbayes}. To do so we suggest testing multiple submodels with different number of nodes and comparing them on the basis of their Bayesian Information Criterion, where  smaller values are preferable \citep[BIC;][]{schwarz1978estimating}. The BIC is available in the \texttt{lme($\cdot$)} output and is known to penalize complex models more strongly than the Akaike Information Criterion \citep[AIC;][]{akaike_new_1974}. Conveniently, the BIC allows a fully Bayesian interpretation \citep{raftery_bayesian_1995}. The difference in BIC between two models is to be interpreted as ``not worth mentioning,'' ``positive,'' ``strong,'' and ``very strong'' evidence against the model with highest value when the difference is between $0$ and $2$, $2$ and $6$, $6$ and $10$, and greater than $10$, respectively. \citep[p.~777]{kass_raftery}. Applications of the spline-based approach within the joint modeling framework can be founded, for example, in \cite{rizopoulos2009fully}, \cite{brown2005flexible}, \cite{ding2008modeling}, and \cite{rizopoulos2011bayesian}.

As explained in \cite{oehlert2014few} and \cite{pinheiro2000mixed}, likelihood ratio tests to compare models \textit{with different fixed effects and same random effects} should be performed using \textit{maximum likelihhod estimation} (ML). However, the default estimation method implemented within the \texttt{lme($\cdot$)} function uses Restricted Maximum Likelihood (REML), because REML, unlike ML, produces unbiased estimates of variance and covariance parameters (the difference in bias vanishes with high number of observations). To meaningfully compare longitudinal submodels in terms of their BIC they should thus be \textit{re-estimated} with ML.

For this illustration, we modeled change in cognitive performance as a function of age in years centered at age 50 (\texttt{Age}), conditioned on age in years at study entry (\texttt{AgeStart}). Thus, we add \texttt{AgeStart} and \texttt{Age} as fixed parameters to the longitudinal submodel. We also modeled \texttt{Age} as a random effect, because visually we might infer differences in rates of decline. Individual observed trajectories of PS are depicted in Figure \ref{mlsc_AD}.

We estimated the following specifications of the longitudinal submodel:
\begin{equation}
\begin{aligned}
\begin{split}
\texttt{lmeFit.mlsc1:}  y_i(t)=& (\beta_0 + b_{0{i}}) + (\beta_1 + b_{1{i}})\text{\texttt{Age}}_i(t)+\beta_4 \text{\texttt{AgeStart}}_i+\epsilon_i(t), \\
\texttt{lmeFit.mlsc2:}  y_i(t)=& (\beta_0 + b_{0{i}}) + (\beta_1 + b_{1{i}})\text{\texttt{Age}}_i(t) + \beta_2 \text{\texttt{Age}}_i(t)^2+\\
&\beta_4 \text{\texttt{AgeStart}}_i+\epsilon_i(t),\\
\texttt{lmeFit.mlsc3:}  y_i(t)=& (\beta_0 + b_{0{i}}) + (\beta_1 + b_{1{i}})\text{\texttt{Age}}_i(t) + \beta_2 \text{\texttt{Age}}_i(t)^2 + \\ &\beta_3 \text{\texttt{Age}}_i(t)^3+\beta_4 \text{\texttt{AgeStart}}_i+\epsilon_i(t),\\
\texttt{lmeFit.mlsc4:}  y_i(t)=& f(\text{\texttt{Age}}_i(t),2)+b_{0{i}} + b_{1{i}}\text{\texttt{Age}}_i(t) +\epsilon_i(t),\\
\texttt{lmeFit.mlsc5:}  y_i(t)=& f(\text{\texttt{Age}}_i(t),3)+b_{0{i}} + b_{1{i}}\text{\texttt{Age}}_i(t) +\epsilon_i(t),\\
\texttt{lmeFit.mlsc6:}  y_i(t)=& f(\text{\texttt{Age}}_i(t),4)+b_{0{i}} + b_{1{i}}\text{\texttt{Age}}_i(t) +\epsilon_i(t).
\end{split}
\end{aligned}
\label{lmeMLSC}
\end{equation}

The first three (\texttt{lmeFit.mlsc1}, \texttt{lmeFit.mlsc2}, \texttt{lmeFit.mlsc3}) estimate, respectively, a polynomial of first, second, and third degree of \texttt{Age}, respectively, with random effects for the intercept and \texttt{Age}. Models 4, 5, 6 (\texttt{lmeFit.mlsc4}, \texttt{lmeFit.mlsc5} \texttt{lmeFit.mlsc6}) estimate cubic spline effects of \texttt{Age} with, respectively, 2, 3 and 4 nodes. Here, too, we included random effects for the intercept and \texttt{Age}. To do so, we used the \texttt{R} code in Section Code 2 lines 3-16 in the Computer Code Section.
%As said previously, longitudinal submodels using splines unfortunately do not reach convergence and therefore will not be discussed further.

We evaluated the adjustments of the six submodels in terms of their respective BIC values. The \texttt{R} code to this end can be founded in Section Code 2 lines 19-25 in the Computer Code Section. The results are displayed in Table \ref{BIC}.

\begin{table}[ht!]
\centering
\begin{tabular}{lrc}
\hline
  & df & BIC \\
\hline
\texttt{m1} &  7 & 29414.87 \\
\texttt{m2} &  8 & 29158.73 \\
\texttt{m3} &  9 & 29167.24 \\
\texttt{m4} &  8 & 29162.86 \\
\texttt{m5} &  9 & 29170.97 \\
\texttt{m6} & 10 & 29178.68 \\
\texttt{m2.1} &  9 & 29155.24 \\
\texttt{m2.2} & 10 & 29163.72 \\
   \hline
\end{tabular}
\caption{BIC values for mixed-effects model comparisons.}
\label{BIC}
\end{table}

For this we concluded that the submodel with a degree-2 polynomial for the fixed effects of \texttt{Age} and with a random intercept and a random \texttt{Age} slope fit the data best (\texttt{m2}, corresponding to \texttt{lmeFit.mlsc2}).

We subsequently expanded this submodel by adding interaction effects between the degree-2 \texttt{Age} polynomial and \texttt{AgeStart}, as shown in Equation \eqref{longexpand}. We again evaluated statistical fit with the BIC. The corresponding \texttt{R} code can be founded in Section Code 3 lines 1-4 in the Computer Code Section.
\begin{equation}
\begin{aligned}
\begin{split}
\texttt{lmeFit.mlsc2.1:} y_i(t)=& (\beta_0 +b_{0{i}}) + (\beta_1 + b_{1{i}})\text{\texttt{Age}}_i(t) + \beta_2 \text{\texttt{Age}}_i(t)^2 + \\ &\beta_3 \text{\texttt{AgeStart}}_i+ \beta_4 \text{\texttt{AgeStart}}_i \times \text{\texttt{Age}}_i(t)+\epsilon_i(t), \\
\texttt{lmeFit.mlsc2.2:} y_i(t)=& (\beta_0 + b_{0{i}}) + (\beta_1 + b_{1{i}})\text{\texttt{Age}}_i(t) + \beta_2 \text{\texttt{Age}}_i(t)^2 + \\ &\beta_3 \text{\texttt{AgeStart}}_i+ \beta_4 \text{\texttt{AgeStart}}_i \times \text{\texttt{Age}}_i(t)\\& +\beta_5 \text{\texttt{AgeStart}}_i \times \text{\texttt{Age}}_i(t)^2+\epsilon_i(t).
\end{split}
\end{aligned}
\label{longexpand}
\end{equation}

The comparison between these two submodels and the previous \texttt{m2} submodel is also displayed in Table \ref{BIC} and indicates that the submodel \texttt{m2.1} (corresponding to \texttt{lmeFit.mlsc2.1}) best fit the data (the \texttt{R} code can be founded in Section Code 3 lines 7-10 in the Computer Code Section).

To conclude this section on the longitudinal submodel of the MLSC PS score, we retain submodel \texttt{lmeFit.mlsc2.1}, which will be joined to the upcoming time-to-event submodel in the final joint model. The function for the true value of PS $\mu_i(t)$ (where the observed PS score $y_i(t) = \mu_i(t) + \epsilon_i(t)$) is
\begin{equation}
\begin{aligned}
\mu_i(t) = &(\beta_0 + b_{0{i}}) + (\beta_1 + b_{1{i}})\text{\texttt{Age}}_i(t) + \beta_2 \text{\texttt{Age}}_i(t)^2 + \beta_3 \text{\texttt{AgeStart}}_i+\beta_4 \text{\texttt{AgeStart}}_i \times \text{\texttt{Age}}_i(t).
\end{aligned}
\label{Longitudinal model_mlsc}
\end{equation}

%An additional point to note is that the model selection procedure presented in this Section may give an unfair advantage to the spline model. This is because the best-fitting spline model is first selected from within a large space of possible spline parameterizations prior to being compared to alternative parametric models. This means that the effective degrees of freedom for the spline model is much larger than for the parametric models and so, all else being equal, the spline model would be selected more often. The addition of penalization of the spline model would improve the longitudinal submodel selection procedure.

\subsubsection{Time-to-event model selection}
\label{survival_mlsc}
As discussed previously, the Cox PH submodel, defined in Equation \eqref{Survival model1}, is often the preferred choice to analyze time-to-event data, given it does not require the user to specify a parametric hazard function. The nonparametric specification of the baseline hazard function within the \texttt{coxph($\cdot$)} function in \texttt{R} uses a B-spline approach, but if desired, regression splines can instead be invoked by setting the \texttt{baseHaz} function argument accordingly. The number and the position of the nodes are automatically selected, but they can also be fixed via the \texttt{lng.in.kn} and the \texttt{knots} function arguments.

In this illustration we specified a nonparametric baseline hazard function within the Cox PH submodel, with factor Smoker (whether or not an individual was a smoker at study entry) as baseline variable, conditioned on age in years at study entry (\texttt{AgeStart}). The corresponding \texttt{R} code can be founded in Section Code 4 lines 1-2 in the Computer Code Section and results are displayed in Table \ref{Cox}. The results show that factor Smoker is significant ($p$-value < 0.001). The estimated coefficient means that being smoker at study entry is associated with a hazard ratio of $\exp(0.519) =  1.680$. That is, being smoker at study entry increases the relative risk of death by $68$\%, compared to a non-smoker at entry.
%Note that a $20$-year difference in age would predict $\exp(20\times 0.03) %= 1.82$, or $82$\% higher relative risk (and not $20 \times 3$\% = $60$\% higher relative risk).
\begin{table}[ht!]
\centering
\begin{tabular}{lccccc}
\hline
&coef & exp(coef) &  se(coef) & z & P\\
\hline
\texttt{Smoker} & 0.519  &1.680 &0.064 & 8.092 & <0.001\\
\texttt{AgeStart}   &0.099  &1.104 &0.004 & -9.815 & <0.001\\
   \hline
\end{tabular}
\caption{Parameter estimates from the Cox PH model.}
\label{Cox}
\end{table}

An important (and often overlooked) assumption of the Cox PH model is that baseline hazard functions for model predictors are proportional \citep{cox1972}. In other words, survival curves for different strata of a given predictor (determined by the particular choices of values for the $\gamma$-variables), have hazard functions that are proportional over time. For example, if the model specifies sex (male vs. female) as a mortality predictor, it is assumed that the hazard functions for males and females are proportional.

The PH assumption can be checked statistically using the Schoenfeld Residuals Test (SRT), which evaluates independence between model residuals and the time variable (here \texttt{Age}). More precisely, SRT checks whether the correlation between the scaled residuals and the time variable differs significantly from zero. If so, the PH assumption is violated, because the residuals are related to time, meaning that the hazard changes in time.

SRT can be applied for each predictor individually and also globally (for the model with all predictors) using the function \texttt{cox.zph($\cdot$)} from the \textbf{survival} package. Table \ref{SRT} displays the results of the SRT and the corresponding \texttt{R} code can be founded in Section Code 4 line 3 in the Computer Code Section.
\begin{table}[ht!]
\centering
\begin{tabular}{lccc}
\hline
& rho & chisq & P \\
\hline
\texttt{Smoker} &-0.068& 8.38 &0.004\\
\texttt{AgeStart}   & 0.061 &6.34& 0.012\\
GLOBAL    &      NA& 15.81 &<0.001\\
   \hline
\end{tabular}
\caption{Schoenfeld Residuals Test form the Cox PH model.}
\label{SRT}
\end{table}

\begin{figure}[!tb]
\centering
%\hspace*{-1in}
\includegraphics[height=9cm,keepaspectratio]{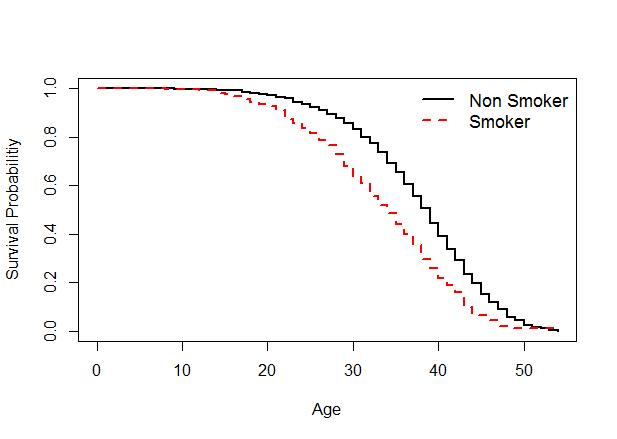}
\caption{Kaplan-Meier estimator of survival probabilities for the two groups smokers versus non smokers.}
\label{KM_MLSC}
\end{figure}

The results of the SRT indicate that the PH assumption does not hold for the factor \texttt{Smoker} and for \texttt{AgeStart} as a predictor ($p$-values $<.05$), nor for the whole submodel ($p$-value $<.05$). When the PH assumption is not met for a predictor of theoretical interest, a stratification procedure for that predictor is required \citep{fox2002cox}. This stratification procedure is supported within the \texttt{coxph($\cdot$)} function inside the \textbf{JMbayes} package. In our particular case, we should thus stratify the time-to-event analysis by smoker status, with two strata, smoker vs non smoker, and also initial age, with an arbitrary number of strata (e.g., 50-59, 60-69, $\ldots$, 91-99 years), where within each the PH assumption is met. For the sake of simplicity within this tutorial, we do not use the stratification procedure and maintain the variables \texttt{Smoker} and \texttt{AgeStart} in the Cox PH submodel, as though the PH assumption were met for both. Kaplan-Meier survival curves by \texttt{Age} by smoker status are depicted in Figure \ref{KM_MLSC}.

\subsubsection{Joint model}
\label{joint_models_mlsc}

We are now ready to explore the association between longitudinal trajectories in processing speed and risk of death. We consider the three association structures as discussed above and, again, use a model comparison approach to select the final model.

\paragraph{The ``current value'' association}
\label{CV_mlsc}
We start with the ``current value'' parameterization defined in Equation \eqref{Survival model_p1}. For our data, this joint model si written as
\begin{equation}
\begin{split}
h_{i}(t) = h_{0}(t) \exp \big(\beta_1 \text{\texttt{Smoker}}_i +\beta_2 \text{\texttt{AgeStart}}_i+ \alpha_1 \mu_i(t)\big),
\end{split}
\label{Survival model_mlsc1}
\end{equation}
with $h_{0}(t)$ denoting the baseline hazard function. We estimated this submodel with the \texttt{R} command displayed in Section Code 5 lines 1-2 in the Computer Code Section. Note that the previously estimated longitudinal and time-to-event submodels are the main arguments of the \texttt{jointModelBayes($\cdot$)} function, and that the ``current value'' is the default association structure. The \texttt{jointModelBayes($\cdot$)} function uses a Metropolis-based Markov chain Monte Carlo (MCMC) algorithm, for which options must be chosen. We discuss these options in the Section Code 5 in the Computer Code Section.

Additionally, in a Bayesian paradigm, prior distributions must be specified for model parameters. A basic approach is to simply select non-informative priors, which is the default for \textbf{JMbayes}. If desired, informative prior distributions can also be specified. For example, in \cite{andrinopoulou2016bayesian}, where the goal was to estimate elaborate models with many predictors in the time-to-event submodel, shrinkage priors were used in order to force small effects to be equal to zero while maintaining true large effects \citep{VANERP201931}. Of importance, \textbf{JMbayes} uses parameter estimates from the individual longitudinal and time-to-event submodels, as illustrated above, as starting values for the MCMC estimation of the joint model. Also, the package relies on a single, long chain, rather than multiple parallel, short chains.

\begin{figure}[!tb]
\hspace{2.5cm}
\begin{minipage}{\textwidth}\raggedright
 \includegraphics[width=300pt]{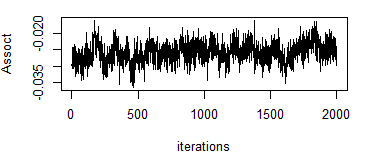}
 \includegraphics[width=300pt]{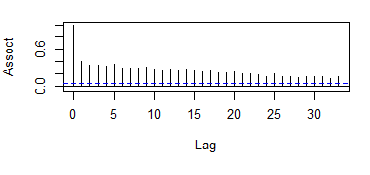}
 \includegraphics[width=300pt]{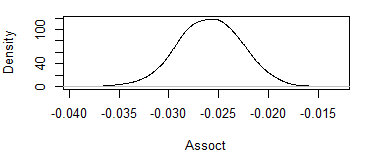}
\end{minipage}
\caption{Trace plot, autocorrelation plot and kernel density plot for the associative parameter \texttt{Assoct} of the joint model with current value parametrization \texttt{jointFit.mlsc1}.}
\label{MCMC_plots}
\end{figure}

As usual, before interpreting the results, we should worry about the estimation quality of the model. Various diagnostic plots are available within \textbf{JMbayes}. The \texttt{plot($\cdot$)} function can be used to visually examine convergence of the \texttt{jointModelBayes($\cdot$)} MCMC estimation. The output includes trace plots, autocorrelation plots, and kernel density estimation plots. An example of these plots for the association parameter fitted within the joint model with current value parametrization can be founded in Figure \ref{MCMC_plots}.

In the trace plots, we seek the so-called ``lazy caterpillar'' pattern, which means that across the various iterations the estimated values of the parameters are within reasonably restricted ranges, and are not suddenly outside such a range. For the autocorrelation plots, which represent the correlations between a solution and that a given lag from it, we seek autocorrelations that become small and as close to zero as quickly as possible (i.e., with small lags), meaning that the solutions of the simulated samples become quickly independent. The kernel density plots are smoothed histograms of the estimated solutions across the simulated samples. We expect these to be unimodal and with small tails. It is also possible to specify MCMC estimation with multiple shorter chains, which can also be diagnosed, but this requires additional \texttt{R} coding: Users must select different initial values within the priors space and then extract the corresponding MCMC output for analysis using the \textbf{coda} MCMC diagnostic package \citep{coda,JMbayes}.

The estimated parameter values can be obtained with the standard \texttt{summary($\cdot$)} function (line 4 of the Section Code 5 in the Computer Code Section). We show the results of this first association structure in Table \ref{CV}. We report the posterior means of all estimated parameters, with their 95\% credible intervals. The \texttt{Assoct} coefficient in the output represents $\alpha_1$ in Equation \eqref{Survival model_mlsc1} and estimates the association between the true processing speed score at time $t$ and the risk of death at that same time. We can interpret this parameter as follows: The percent reductions in mortality risk associated with the current true value of processing speed for women is calculated as $1 - \text{HR} \times 100$, where HR is the hazard ratio corresponding to a one-unit difference in the current true value of processing speed $\mu_i(t)$. Thus, HR is calculated as $\exp(-0.026)=0.974$, which gives $1-0.974=0.026$, and translates to $2.6\%$ lower risk of death per one unit increase in the current true value of processing speed. The coefficient for smoker statuts (0.445) indicates a strong association with mortality, with a $\exp(0.445)$ = $ 1.560$-fold increase in risk of death in smokers, compared to non smokers.

\begin{table}[ht]
\centering
\begin{tabular}{lrrr}
  \hline
  & Value & 2.5\% & 97.5\% \\
  \hline
\texttt{Smoker} & 0.445 & 0.328 & 0.555\\
\texttt{AgeStart} & -0.048 &-0.055 & -0.038\\
\texttt{Assoct} & -0.026 &-0.032 & -0.020\\
\texttt{tauBs} &  12.186 & 4.643 & 25.541\\ \hline

\texttt{(Intercept)}  &    8.806 & 6.506 &11.060\\
\texttt{Age}       &    0.192 & 0.079 & 0.302\\
\texttt{AgeStart}     &   -0.054 &-0.090 &-0.018\\
I(\texttt{Age}$^2$)   &  -0.007 &-0.007 &-0.006\\
\texttt{Age:AgeStart} & -0.005 &-0.006 &-0.003\\
\texttt{$\sigma$} &  0.946 & 0.916 & 0.977\\
\texttt{D[1, 1]} & 31.550 & 28.805 & 34.265\\
\texttt{D[2, 1]} & -0.320 & -0.402 & -0.229 \\
\texttt{D[2, 2]} & 0.078 & 0.073 & 0.084 \\
   \hline
\end{tabular}
\caption{Parameter estimates and 95\% credible intervals for the ``current value'' association. D[i, j] denotes the $ij$-element of the covariance matrix of the random effects. tauBs is a hyperparameter relative to the distribution of the baseline hazard function $h_0 (t)$ and is typically not interpreted.}
\label{CV}
\end{table}

%%%%%%%%%%%%%%%%%%%%%%%%%%%%%%%%%%%%%%%%%%%%%%%%%%%%%
%DONE
%what this means in terms of the actual variable, not just the abstract model parameter
%%%%%%%%%%%%%%%%%%%%%%%%%%%%%%%%%%%%%%%%%%%%%%%%%%%%%

\paragraph{The ``current value plus slope'' association}
\label{CVprocessing speed_mlsc}

This second association structure takes the form
\begin{equation}
\begin{split}
h_{i}(t) = h_{0}(t) \exp \big(\beta_1 \text{\texttt{Smoker}}_i +\beta_2 \text{\texttt{AgeStart}}_i + \alpha_1 \mu_i(t)+\alpha_2 \mu_i^{'}(t)\big),
\end{split}
\label{CVS}
\end{equation}
and specifies that the hazard of experiencing the event at time $t$ is now assumed to be associated both with the true value of processing speed at time $t$, $\mu_i(t)$, and its current rate of change (i.e., slope) at time $t$, noted $\mu_i^{'}(t)$.

Recalling Equation \eqref{Longitudinal model_mlsc}, the temporal variable is \texttt{Age}. The derivative of Equation \eqref{Longitudinal model_mlsc} with respect to the temporal variable \texttt{Age} can be written as
\begin{equation}
\begin{aligned}
\mu_i^{'}(t)&=\frac{ \partial \mu_i(t)}{\partial \text{\texttt{Age}}}\\
&=\beta_1 + b_{1{i}} +  \beta_2 (2 \times \text{\texttt{Age}}_i(t))  +\beta_4 \text{\texttt{AgeStart}}_i.
\end{aligned}
\label{derivative_MLSC}
\end{equation}
The \texttt{R} code for specifying the derivative form of this linear submodel and for running the corresponding joint model can be found in the Computer Code Subsections 6 and 7.

We show the results of the ``current value plus slope'' parametrization in Table \ref{CVS}.
\begin{table}[ht]
\centering
\begin{tabular}{lrrrrr}
\hline
  & Value & 2.5\% & 97.5\% \\
  \hline
\texttt{Smoker} & 0.424 & 0.315 & 0.573\\
\texttt{AgeStart} & -0.056 & -0.066 & -0.048\\
\texttt{Assoct} & 0.001 & -0.008 & 0.010\\
\texttt{AssoctE} & -1.600 & -1.984 & -1.243 \\
\texttt{tauBs} &14.250 & 5.747 & 28.615 \\ \hline

\texttt{(Intercept)} & 8.502 & 6.232 & 10.765\\
\texttt{Age} & 0.185 & 0.070 & 0.295\\
\texttt{AgeStart} & -0.049 & -0.086 & -0.012\\
I(\texttt{Age}$^2$) &  -0.007 & -0.007 & -0.006\\
\texttt{Age:AgeStart} &-0.005 & -0.006 & -0.003 \\
\texttt{$\sigma$} &0.946 & 0.910 & 0.981  \\
\texttt{D[1, 1]} & 31.093 & 28.992 & 33.549\\
\texttt{D[2, 1]} & -0.301 & -0.396 & -0.216\\
\texttt{D[2, 2]} & 0.079 & 0.073 & 0.085\\
   \hline
\end{tabular}
\caption{Parameter estimates and 95\% credible intervals for the ``current value plus slope'' association. D[i, j] denotes the $ij$-element of the covariance matrix of the random effects. tauBs is a hyperparameter relative to the distribution of the baseline hazard function $h_0 (t)$ and is typically not interpreted.}
\label{CVS}
\end{table}
\texttt{Assoct} represents $\alpha_1$ and \texttt{AssoctE} represents $\alpha_2$ in Equation \eqref{Survival model_p2}, and quantify, respectively, the association between the current true value and the current rate of change of processing speed and the relative risk of death. We obtain $\alpha_1=0.001$ and $\alpha_2=-1.600$, with the 95\% credible interval of \texttt{Assoct} that contains 0, while that of \texttt{AssoctE} does not. Clearly, then, the rate of change in perceptual speed at a given time point is a much more important predictor of  than the actual true value of speed at that same time point. The lower the rate of change, the greater the odds of dying.

\begin{figure}[!tb]
\centering
\hspace*{0.5in}
\includegraphics[height=9cm,keepaspectratio]{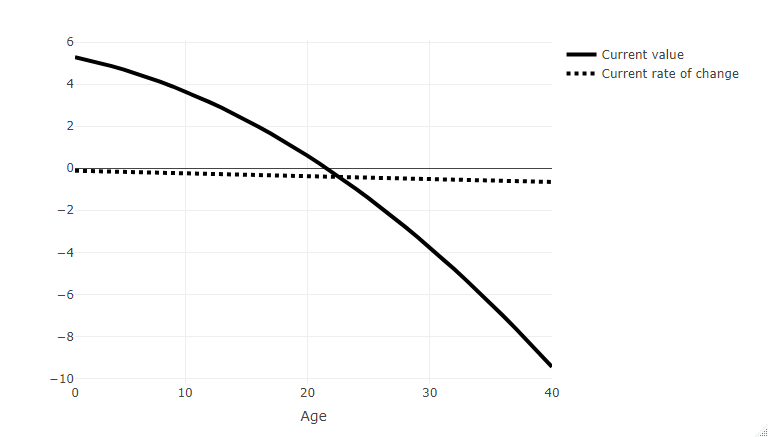}
\caption{Average current value and average current rate of change (calculated without the random effects) for \texttt{AgeStart} fixed at its median value ($63$).}
\label{CV_CVS}
\end{figure}

To interpret the effect of \texttt{AssoctE} on the risk of death, we could proceed as usual, by relating it to a one-unit change. However, whereas a one-unit value is a meaningful comparison for the current true value of speed, it is excessively large for the rate of change. To show this, we plotted in Figure \ref{CV_CVS} the current true value $\mu(t)$ estimated from the previous joint model with the ``current value'' association (continuous line) and the current rate of change $\mu^{'}(t)$ (dashed line) from the joint model ``current value plus slope'', as a function of the \texttt{Age} variable. We can immediately see that whereas the current value changes over a large range (from almost 6 to -10), the rate of change varies much less (from about 0 to less than -1). Hence, whereas a one-unit change could be useful in interpreting the former effect, it is less so for the latter effect (it would result in a $(1-\exp(-1.600)) \times 100=79.81\%$ reduction in mortality risk, which is simply huge). Coming back to Figure \ref{CV_CVS}, we note that over an interval of $40$ years, the current true value decreases by $14.722$ points, whereas the rate of change decreases by only $0.535$ points. This translates to an average decrease over a 10-year period of $3.684$ for the current true value and $ 0.134$ for the rate of change, with associated percent reductions in mortality risk of $ 9.13\%$ (calculated as $(1-\exp(-0.026 \times 3.684)) \times 100$) and $19.3\%$ (calculated as $(1-\exp(-1.6 \times 0.133)) \times 100$), respectively. We believe this exercice to be most useful to properly interpret the effect sizes stemming from this model, because, typically, the range of the current true value is much larger than that of its rate of change.

%---------------------------------------------------------------------------------------------
%---------------------------------------------------------------------------------------------
%---------------------------------------------------------------------------------------------

\paragraph{The ``shared random effects'' association}
\label{SRE_mlsc}

The third joint model can be written as
\begin{equation}
\begin{split}
h_{i}(t) = h_{0}(t) \exp \big(\beta_1 \text{\texttt{Smoker}}_i +\beta_2 \text{\texttt{AgeStart}}_i
 + \alpha_3 b_{0_i} + \alpha_4 b_{1_{i}}\big),
\end{split}
\label{Survival model_mlsc3}
\end{equation}
where $\alpha_3$ and $\alpha_4$ are the parameters of association of $b_{0i}$ and $b_{1i}$, which are defined in Equation \eqref{lmeMLSC} as the individual deviations from the sample average intercept and slope values, respectively. $\alpha_3$ and $\alpha_4$ define the association between the random intercept and the random slope effects, respectively, and the hazard of death. More precisely, for individuals having the same deviation from the average slope, the hazard ratio for a one-unit increase in individual deviation from the average intercept is $\exp (\alpha_3)$. Conversely, for individuals having the same deviation from the average intercept, the hazard ratio for a one unit increase in individual deviation from the average slope is $\exp (\alpha_4)$. The code for estimating the joint model with the ``shared random effects'' parameterization can be founded in Section Code 8 lines 1-3 in the Computer Code Section.

The results are displayed in Table \ref{SRE}. \texttt{Assoct:(Intercept)} represents $\alpha_3$ and \texttt{Assoct:Age} represents $\alpha_4$ in Equation \eqref{Survival model_mlsc3}.
Subject-specific deviations from the sample average intercept are not predictive of the risk of death, whereas subject-specific deviations from the sample average slope are. To interpret this latter effect properly, rather than relating the reduction in mortality risk to a one-unit change, we relate it to a one-SD change, where D[2, 2]=0.080. This risk reduction is calculated as $ 1 - \exp(\alpha_4 \times \text{std}(b_1)) = 1 - \exp( -1.641 \times \sqrt{0.080}) = 0.371$. Thus, for a 1SD decrease in slope, the percent increase in mortality risk is $37.1\%$. Participants with greater cognitive decline have greater odds of dying than those declining less.

\begin{table}[ht]
\centering
\begin{tabular}{lrrrrr}
\hline
& Value & 2.5\% & 97.5\%\\
\hline
\texttt{Smoker} & 0.410 & 0.291 & 0.536\\
\texttt{AgeStart} & -0.048 & -0.056 & -0.041\\
\texttt{Assoct:(Intercept)} & 0.004 & -0.006 & 0.014\\
\texttt{Assoct:Age} &  -1.641 & -1.884 & -1.397\\
\texttt{tauBs} & 7.426 & 2.792 & 16.424\\ \hline

\texttt{(Intercept)} & 8.330 & 6.050 & 10.515\\
\texttt{Age} & 0.192 & 0.079 & 0.308\\
\texttt{AgeStart} & -0.046 & -0.081 & -0.009\\
I(\texttt{Age}$^2$) & -0.007 & -0.007 & -0.007\\
\texttt{Age:AgeStart} & -0.005 & -0.007 & -0.003  &&\\
$\sigma$ & 0.957 & 0.924 & 0.993\\
\texttt{D[1, 1]} & 31.438 & 29.167 & 33.827 \\
\texttt{D[2, 1]} &-0.322 & -0.411 & -0.224 \\
\texttt{D[2, 2]} & 0.080 & 0.074 & 0.085 \\
   \hline
\end{tabular}
\caption{Parameter estimates and 95\% credible intervals for the ``shared random effects'' association. D[i, j] denotes the $ij$-element of the covariance matrix of the random effects. tauBs is a hyperparameter relative to the distribution of the baseline hazard function $h_0 (t)$ and is typically not interpreted.}
\label{SRE}
\end{table}

\paragraph{Joint model with interaction}
\label{interaction_mlsc}
We have illustrated the three association structures discussed previously in the joint models. Before reaching a final substantive conclusion about our data set, we would like to illustrate an additional extension of the association function $f(\cdot)$ in Equation \eqref{Survival_model_general}. We illustrate interactions of predictors with the current true value $\mu_i(t)$ and/or current rate of change $\mu_i^{'}(t)$ terms.

We illustrate this by extending the model \texttt{jointFit.mslc1} to include an \textit{interaction term} between the current true value of processing speed $\mu_i(t)$ and the smoker status. The corresponding time-to-event submodel can be written as
\begin{equation}
\begin{split}
\begin{aligned}
h_{i}(t) = &h_{0}(t) \exp \big(\beta_1 \text{\texttt{Smoker}}_i +\beta_2 \text{\texttt{AgeStart}}_i + \alpha_1 \mu_i(t)+ \alpha_5 (\mu_i(t) \times \text{\texttt{Smoker}}_i) \big),
\end{aligned}
\end{split}
\label{Survival model_mlsc4}
\end{equation}
where the interaction coefficient $\alpha_5$ expresses the difference in mortality prediction of $\mu_i(t)$ between smokers (\texttt{Smoker} = 1) compared to non-smokers (\texttt{Smoker} = 0).

The interaction is specified within the \textbf{JMbayes} package using the \texttt{transFun} argument, which takes as its value a generic function \texttt{tf1} that, in turns, takes arguments \texttt{x} for the longitudinal parameter of interest and \texttt{data} for the predictor with which the longitudinal parameter interacts. This \texttt{tf1} function has to be saved under its generic name (\texttt{tf1}) and defined in the workspace \citep[see][for more complex examples, such as multiple transformation functions applied simultaneously to multiple longitudinal parameters]{JMbayes}. The \texttt{R} code allowing one to write the generic function \texttt{tf1} related to the submodel with interaction in Equation \eqref {Survival model_mlsc4} can be founded in Section Code 9 line 1 in the Computer Code Section. The \texttt{R} code allowing one to \textit{source} the function \texttt{tf1} and to subsequently estimate the joint model with interaction can be founded in Section Code 9 lines 2-3 in the Computer Code Section.

The results are presented in Table \ref{INT}. The coefficient \texttt{Smoker} means that being smoker (versus non-smoker) increases by 64.54\% $(\exp(0.498)-1) \times 100)$ the risk of dying. \texttt{Assoct} ($\alpha_1$ in equation \eqref{Survival model_mlsc4}) represents the association between the current true value of processing speed and the mortality risk and \texttt{Assoct:Smoker} ($\alpha_5$ in equation \eqref{Survival model_mlsc4}) represents the \textit{interaction effect} between the current true value of processing speed and the factor \texttt{Smoker} on the mortality risk. These results indicate that the current value of processing speed $\mu_i(t)$ is negatively associated with the risk of death (\texttt{Assoct} $=-0.028$). Going back to the previous interpretation of the \texttt{Assoct} effect of $3.684$ units over a 10-year period, rather than a one-unit change (cf. Figure \ref{CV_CVS}), we obtain a $9.8\%$ (calculated as $(1-\exp(-0.028 \times 3.684)) \times 100$) increase in mortality risk. Note that this effect is very similar to the associative effect previously estimated for the ``current value'' parametrization, and remains different from zero. The interaction effect \texttt{Assoct:Smoker} is unimportant ($95\%$ credible interval contains 0), meaning that the effect of the current true value of processing speed on the mortality risk does not differ between smokers and non-smokers.

\begin{table}[ht]
\centering
\begin{tabular}{lrrrrr}
  \hline
  & Value & 2.5\% & 97.5\% \\
  \hline
 \texttt{Smoker} &  0.498 & 0.306 & 0.667\\
 \texttt{AgeStart} & -0.049 & -0.057 & -0.043\\
 \texttt{Assoct} & -0.028 & -0.034 & -0.021\\
 \texttt{Assoct:Smoker} & 0.007 & -0.007 & 0.020\\
 \texttt{tauBs} &6.989 & 2.420 & 18.851 \\ \hline

 \texttt{(Intercept)} & 8.746 & 6.453 & 11.048 \\
  \texttt{Age} & 0.191 & 0.079 & 0.300\\
  \texttt{AgeStart} & -0.054 & -0.090 & -0.016\\
  I(\texttt{Age}$^2$) & -0.007 & -0.007 & -0.006\\
  \texttt{Age:AgeStart} & -0.005 & -0.006 & -0.003 \\
  $\sigma$ & 0.938 & 0.906 & 0.968 \\
  \texttt{D[1, 1]} & 32.501 & 30.020 & 35.273\\
  \texttt{D[2, 1]} & -0.346 & -0.453 & -0.241  \\
  \texttt{D[2, 2]} & 0.079 & 0.073 & 0.084 \\
   \hline
\end{tabular}
\caption{Parameter estimates and 95\% credible intervals for the joint model with interaction. D[i, j] denotes the $ij$-element of the covariance matrix of the random effects. tauBs is a hyperparameter relative to the distribution of the baseline hazard function $h_0 (t)$ and is typically not interpreted.}
\label{INT}
\end{table}

\paragraph{Model comparison}
\label{comparison_mlsc}

We can finally proceed to compare statistically the various association structures of the estimated joint models. To do so, we rely on the Deviance Information Criterion \citep[DIC;][]{Spiegelhalter}, with smaller values indicating better model adjustments to the data. The \texttt{R} code computing the joint models comparison can be founded in Section Code 10 line 1 in the Computer Code Section. The results of joint models comparison can be founded in Table \ref{DIC}.

\begin{table}[ht]
\centering
\begin{tabular}{lrrrr}
  \hline
     & df   &   LPML    &  DIC   &    pD \\
  \hline
\texttt{jointFit.mlsc1} & 5591 & -26425.21 & 51793.14 & 5199.108\\
\texttt{jointFit.mlsc2} & 5592 & -26333.20 & 51606.23 & 5186.702\\
\texttt{jointFit.mlsc3} & 5592 & -26350.51 & 51791.15 & 5229.485\\
\texttt{jointFit.mlsc4} & 5592 & -26357.66 & 51773.84 & 5205.650 \\
   \hline
\end{tabular}
\caption{Join models comparison.}
\label{DIC}
\end{table}

LPML is the log-pseudo-marginal-likelihood value, DIC is the deviance information criterion and pD refers to the effective number of model parameters. Based on the DIC values, the evidence is strong to prefer the \texttt{jointFit.mlsc2} model, which includes the current true value and the current rate of change of processing speed as predictors of the risk of death. Recall that within this model, the effect of the current true value in speed on the hazard of dying was not different from zero, whereas the association between the rate of change in speed and survival was strong. More precisely, individuals experiencing a greater rate of change, equivalent to what is expected over a 10-year interval, have a 19.3\% greater mortality risk than those with shallower rates of change. Thus, we can conclude that cognitive performance per se, assessed via perceptual speed, does not relate strongly to mortality. Rather, it is the rate of change in cognitive performance that is predictive of survival, as hypothesized by the terminal decline and terminal drop hypotheses.

\section{Discussion}
\label{discussion}

In this tutorial, we presented an overview of the joint longitudinal and time-to-event modeling framework, presented a comparison of various \texttt{R} packages allowing this type of analysis, and illustrated how to use the most comprehensive of these \texttt{R} packages, \textbf{JMbayes}, to estimate a series of joint models. We illustrated joint models with real data from a psychological study on adult cognitive development and provided guidelines to choose among a series of theoretically motivated alternative specifications of joint models.

Due to space constraints and to limit complexity, we have not addressed other advanced features of the \textbf{JMbayes} package. These include functionalities for dynamic predictions, where survival probabilities can be estimated for a individual $i$ that was not part of the estimation sample, and for less common association structures. For instance, the package allows the ``lagged effects'' parametrization, where the hazard of experiencing the event at time $t$ is associated with the level of the longitudinal measure at a previous time point $t-p$, hence with a $p-$lag. For example, we might want to study how cognitive development during childhood might affect mortality later on in life. Thus, the longitudinal measures representing cognitive development during childhood at time $t-p$ could be related to mortality at time $t$, with an important (probably around 50-so years) lag $p$. An other association structure implemented in \textbf{JMbayes} is the ``cumulative effects'' parametrization, whereby the hazard of experiencing the event at time $t$ is associated with the whole information relative to the trajectory of the longitudinal measure up to time $t$. That is, rather than relating the value or slope on the true trajectory to the event, the whole surface under the trajectory constitutes the predictive information. For example, we might want to relate how learning characteristics during a critical period relate to mortality. In this case, performance during a given learning trial might be less informative or mortality than the entire learning history, as represented by the area under the learning curve.

%%%%%%%%%%%%%%%%%%%%%%%%%%%%%%%%%%%%%%%%%%%%%%%%%%%%%%%%%%%%%%%%%%%%%%
%DONE
% this is where I would mention the specification of interactions, and point readers to the above example as supplemental materials...
%%%%%%%%%%%%%%%%%%%%%%%%%%%%%%%%%%%%%%%%%%%%%%%%%%%%%%%%%%%%%%%%%%%%%%
Recently, \textbf{JMbayes} was extended to support simultaneous modeling of multiple longitudinal outcomes (using the function \texttt{mvJointModelBayes($\cdot$)}). This multiple parametrization may be useful when several longitudinal measures are recorded and may jointly impact the risk of death. For instance if several measure of cognitive development are recorded \citep[as processing speed, fluid intelligence and crystallized intelligence in][]{Rabbitt2004}, they may jointly impact the risk of death. We also encourage interested readers to explore the developments of other \texttt{R} packages and other software allowing for joint model estimation.

The primary objective of this tutorial was to illustrate joint modeling of longitudinal and time-to-event data using the \textbf{JMbayes} package. Rather than reviewing the technical details of the joint modeling methodology, we aimed instead at providing researchers in behavioral sciences a practical understanding of what this powerful methodology can offer in substantive terms and how to apply it to their data. We hope this information will encourage adoption of the joint longitudinal-time-to-event modeling framework in research domains where it is currently underutilized and that should benefit from it.

\clearpage

\section{Computer Code}

\subsection{Code 1}
\label{Code1}

\begin{linenumbers}
\begin{knitrout}\small
\begin{alltt}
install.packages("JMbayes")
install.packages("rjags")
library(JMbayes)
library(rjags)
\end{alltt}
\end{knitrout}
\end{linenumbers}

\subsection{Code 2}
\label{Code 2}

\begin{linenumbers}
\linenumbers[1]
\begin{knitrout}\small
\begin{alltt}
ctrl <- lmeControl(opt="optim")

lmeFit.mlsc1 <- lme(FS_SPD ~  Age_50+AgeStart,
                    data = MLSC_fem2, random = ~ Age_50 | ID)
lmeFit.mlsc2 <- lme(FS_SPD ~  Age_50+I(Age_50^2)+AgeStart,
                    data = MLSC_fem2, random = ~ Age_50 | ID)
lmeFit.mlsc3 <- lme(FS_SPD ~  Age_50+I(Age_50^2)+I(Age_50^3)+AgeStart,
                    data = MLSC_fem2, random = ~ Age_50 | ID)


lmeFit.mlsc4 <- lme(FS_SPD ~  ns(Age_50,2)+AgeStart, data = MLSC_fem2,
                     random = ~ Age_50 | ID)
lmeFit.mlsc5 <- lme(FS_SPD ~  ns(Age_50,3)+AgeStart, data = MLSC_fem2,
                     random = ~ Age_50 | ID)
lmeFit.mlsc6 <- lme(FS_SPD ~  ns(Age_50,4)+AgeStart, data = MLSC_fem2,
                     random = ~ Age_50 | ID)


m1<-update(lmeFit.mlsc1, method = "ML" )
m2<-update(lmeFit.mlsc2, method = "ML" )
m3<-update(lmeFit.mlsc3, method = "ML" )

m4<-update(lmeFit.mlsc4, method = "ML" )
m5<-update(lmeFit.mlsc5, method = "ML" )
m6<-update(lmeFit.mlsc6, method = "ML" )

BIC(m1,m2,m3,m4,m5,m6)
\end{alltt}
\end{knitrout}
\end{linenumbers}
\texttt{ctrl <- lmeControl(opt="optim")} changes the optimizer to be used for \texttt{lme} function (\texttt{nlminb} is the default).

\subsection{Code 3}
\label{Code 3}

\begin{linenumbers}
\linenumbers[1]
\begin{knitrout}\small
\begin{alltt}
lmeFit.mlsc2.1 <- lme(FS_SPD ~  Age_50*AgeStart+I(Age_50^2),
                      data = MLSC_fem2, random = ~ Age_50 | ID)
lmeFit.mlsc2.2 <- lme(FS_SPD ~  Age_50*AgeStart+I(Age_50^2)*AgeStart,
                      data = MLSC_fem2, random = ~ Age_50 | ID)


m2.1<-update(lmeFit.mlsc2.1, method = "ML" )
m2.2<-update(lmeFit.mlsc2.2, method = "ML" )

BIC(m2,m2.1,m2.2)
\end{alltt}
\end{knitrout}
\end{linenumbers}

\subsection{Code 4}
\label{Code 4}

\begin{knitrout}\small
\begin{linenumbers}
\linenumbers[1]
\begin{alltt}
Coxfit_fem <- coxph(Surv(AgeLastObserved_2012_50, Dead_By2012) ~ Smoker+AgeStart,
                    data = MLSC_ID_fem2,x=TRUE)
print(cox.zph(Coxfit_fem))
\end{alltt}
\end{linenumbers}
\end{knitrout}

In the joint modeling context we need to set \texttt{x = TRUE} (or equivalently \texttt{model = TRUE}) in the call of the \texttt{coxph($\cdot$)} function in order that the design matrix used in the Cox model is returned in the object fit (required for the joint model). Importantly, note that the scale of the temporal variable (here \texttt{Age} and \texttt{AgeLastObserved\_2012\_50}) should be identical in the longitudinal and in the survival models.
The indicator for the event (alive or dead) is \texttt{Dead\_By2012}.

\subsection{Code 5}
\label{Code 5}

\begin{knitrout}\small
\begin{linenumbers}
\linenumbers[1]
\begin{alltt}
jointFit.mlsc1 <- jointModelBayes(lmeFit.mlsc2.1, Coxfit_fem,
                                  timeVar = "Age_50",n.iter = 30000)

summary(jointFit.mlsc1)

Call:
jointModelBayes(lmeObject = lmeFit.mlsc2.1, survObject = Coxfit_fem,
    timeVar = "Age_50", n.iter = 30000)

Data Descriptives:
Longitudinal Process		Event Process
Number of Observations: 5517	Number of Events: 1858 (66.8%)
Number of subjects: 2780

Joint Model Summary:
Longitudinal Process: Linear mixed-effects model
Event Process: Relative risk model with penalized-spline-approximated
		baseline risk function
Parameterization: Time-dependent value

      LPML      DIC       pD
 -26425.21 51793.14 5199.108

Variance Components:
             StdDev    Corr
(Intercept)  5.6169  (Intr)
Age_50       0.2801 -0.2035
Residual     0.9464

Coefficients:
Longitudinal Process
                  Value Std.Err Std.Dev    2.5%   97.5%      P
(Intercept)      8.8057  0.1290  1.1708  6.5061 11.0604 <0.001
Age_50           0.1922  0.0040  0.0574  0.0786  0.3024  0.001
AgeStart        -0.0542  0.0023  0.0189 -0.0899 -0.0176  0.002
I(Age_50^2)     -0.0066  0.0000  0.0002 -0.0069 -0.0063 <0.001
Age_50:AgeStart -0.0048  0.0001  0.0009 -0.0065 -0.0030 <0.001

Event Process
             Value Std.Err Std.Dev    2.5%   97.5%      P
SmokerTRUE  0.4449  0.0077  0.0567  0.3280  0.5549 <0.001
AgeStart   -0.0480  0.0010  0.0044 -0.0552 -0.0380 <0.001
Assoct     -0.0259  0.0003  0.0031 -0.0321 -0.0199 <0.001
tauBs      12.1863  1.2287  5.4554  4.6427 25.5413     NA

MCMC summary:
iterations: 30000
adapt: 3000
burn-in: 3000
thinning: 15
time: 17.6 min
\end{alltt}
\end{linenumbers}
\end{knitrout}

\texttt{lmeFit.mlsc2.1} is the longitudinal submodel, \texttt{Coxfit\_fem} is the survival Cox submodel and \texttt{timeVar = "Age\_50"} means that the temporal variable is \texttt{Age\_50}. Note that no specification of the association structure is necessary, because the ``current value'' parameterization is the default implementation. Here we specified that, after the burn-in period, the MCMC algorithm should run for $30000$ iterations.

\subsection{Code 6}
\label{Code 6}

\begin{knitrout}\small
\begin{linenumbers}
\linenumbers[1]
\begin{alltt}
dForm <- list(fixed= ~1 + I(2*Age_50) +AgeStart ,
              random = ~ 1, indFixed = c(2,4,5), indRandom = 2)
\end{alltt}
\end{linenumbers}
\end{knitrout}

This \texttt{R} code means that the derivative of the current value ($\mu_i(t)$), with respect to \texttt{Age\_50}, is equal for the fixed part (\texttt{fixed=}) to an intercept plus $2$ times the \texttt{Age\_50} variable plus \texttt{AgeStart} ($\beta_1 +  \beta_2 (2 \times \text{Age}_i(t)) +\beta_4 \text{AgeStart}$ in Equation (13)).
The derivative of the random part of the model in Equation (13), with respect to \texttt{Age\_50}, equals an intercept ($b_1$ in Equation (13)).

Arguments \texttt{indFixed=c(2,4,5)} and \texttt{indRandom=2} refer to indices of the fixed and random model parameters corresponding to the retained (non-zero) arguments in the derivative of Equation (13). Thus, for the fixed parameters, indices \texttt{c(2,4,5)} relate to row numbers in the output of \texttt{summary(jointFit.mslc1)} under \texttt{Longitudinal Process} that correspond to the appropriate fixed model parameters  $\{\beta_1,\beta_2,\beta_4\}$ (Section Code 5, lines 31-37). Similarly, for the random parameters, index 2 relates to row number relative to the random parameter $b_{1_{i}}$, found under \texttt{Variance Components} (Section Code 5, lines 24-28).

\subsection{Code 7}
\label{Code 7}

\begin{knitrout}\small
\begin{linenumbers}
\linenumbers[1]
\begin{alltt}
jointFit.mlsc2 <- update(jointFit.mlsc1, param = "td-both", extraForm = dForm)
\end{alltt}
\end{linenumbers}
\end{knitrout}

\subsection{Code 8}
\label{Code 8}

\begin{knitrout}\small
\begin{linenumbers}
\linenumbers[1]
\begin{alltt}
jointFit.mlsc3<- jointModelBayes(lmeFit.mlsc2.1, Coxfit_fem,
                                 timeVar = "Age_50",param = "shared-RE",
                                 n.iter = 30000)
\end{alltt}
\end{linenumbers}
\end{knitrout}

\subsection{Code 9}
\label{code 9}

\begin{knitrout}\small
\begin{linenumbers}
\linenumbers[1]
\begin{alltt}
tf1<-function (x, data) {cbind(x, "Smoker" = x * (data\$Smoker=='TRUE'))}
jointFit.mlsc4 <- update(jointFit.mlsc1, transFun = tf1)
\end{alltt}
\end{linenumbers}
\end{knitrout}

This \texttt{tf1} function has to be saved under its generic name (\texttt{tf1}) in a \textit{separate} \texttt{R} file and defined in the workspace. In this code, \texttt{jointFit.mlsc4} calls \texttt{jointFit.mlsc1} (the joint model with the current value parametrization) and the function \texttt{tf1}. It means that \texttt{jointFit.mlsc4} will look for arguments \texttt{x} and \texttt{data} in the previously estimated model \texttt{jointFit.mlsc1}. \texttt{x} is the association structure used in \texttt{jointFit.mlsc1}  (here the current value $\mu_i(t)$) and \texttt{data} are the data used in the estimation of \texttt{jointFit.mlsc1}. Then, the function \texttt{\{cbind(x, "Smoker" = x * (data\$Smoker=='TRUE'))\}} defines the columns of the design matrix $X_i(t)$ relative to the association structure (see Equation (1)). Here \texttt{\{cbind(x, "Smoker" = x * (data\$Smoker=='TRUE'))\}} means that we will estimate first an simple effect of $\mu_i(t)$ (\texttt{x}), and then an interaction between the current value $\mu_i(t)$ and the factor \texttt{Smoker} (\texttt{"Smoker" = x * (data\$Smoker=='TRUE'))}).

\subsection{Code 10}
\label{code 10}

\begin{knitrout}\small
\begin{linenumbers}
\linenumbers[1]
\begin{alltt}
anova(jointFit.mlsc1,jointFit.mlsc2,jointFit.mlsc3,jointFit.mlsc4)
\end{alltt}
\end{linenumbers}
\end{knitrout}

\section*{Acknowledgements}
The authors gratefully acknowledge \textit{Lifebrain, Grant number 732592 - H2020-SC1-2016-2017/H2020-SC1-2016-RTD}.

\bibliographystyle{apacite}
\bibliography{Tutorial_PM_PG}

\begin{table}[!hb]
\centering
\setcellgapes{1pt}
\makegapedcells
\newcolumntype{L}[1]{>{\raggedright\arraybackslash }b{#1}}
\newcolumntype{C}[1]{>{\centering\arraybackslash }b{#1}}
%\begin{tabularx}{16cm}{L{4cm}|L{12cm}}
\begin{tabular}{L{3cm}|L{13cm}}

 \multicolumn{2}{c} {JM \citep{JM,rizopoulos_book}} \\ \hline \hline

   Estimation method &  Frequentist approach with maximum likelihood estimation using an em algorithm   \\ \hline
   Association structure & Current value, current value and slope, shared 		random effects, lagged effects, cumulative effects and free-form association structures implemented  \\ \hline
   Longitudinal submodel & Longitudinal submodel fitted with the \texttt{nlme($\cdot$)} function from the \textbf{nlme} package. Allows a univariate normally distributed response and non linear effects of covariates fitted with spline \\ \hline
  Time-to-event submodel & Time-to-event submodel fitted with the \texttt{coxph}($\cdot$) function from the \textbf{survival} package. A number of relative risk and accelerated failure time survival model options are available, including Weibull, piecewise proportional hazards, Cox proportional hazards, and proportional hazards with a spline-approximated baseline risk function. A single time-to-event outcome or competing risks are feasible. The time-to-event submodel supported the stratification procedure\\ \hline
   Model comparison & Joint models comparison can be performed with a Likelihood Ratio Test (LRT) implemented with the \texttt{anova}($\cdot$) function\\ \hline
   Post-fit analysis & Various post-fit functions including goodness-of-fit analyses, plots, predicted trajectories, individual dynamic prediction of the event and predictive accuracy assessment are available\\ \hline \\

\end{tabular}
\caption{Main functionalities of \texttt{R} package \textbf{JM}.}
\label{Table JM}
\end{table}

%---------------------------------------------------------------------------

%--------------------------------------------------------------------

\begin{table}[!hb]
\centering
\setcellgapes{1pt}
\makegapedcells
\newcolumntype{L}[1]{>{\raggedright\arraybackslash }b{#1}}
\newcolumntype{C}[1]{>{\centering\arraybackslash }b{#1}}
%\begin{tabularx}{16cm}{L{4cm}|L{12cm}}
\begin{tabular}{L{2.2cm}|L{15cm}}

   \multicolumn{2}{c} {JMbayes \citep{JMbayes}} \\ \hline \hline

   Estimation method &  Bayesian approach using a MCMC algorithm. A long chain methodology is implemented. Multiple chains are also feasible with a little programming \\ \hline
   Association structure & Current value, current value and slope, shared random effects, lagged effects, cumulative effects and free-form association structure are implemented \\ \hline
   Longitudinal submodel & Continuous and categorical longitudinal outcomes are allowed. Longitudinal submodel fitted with the \texttt{nlme($\cdot$)} function from the \textbf{nlme} package. Allows one to fit non linear effects of covariates with spline. A very recent extension of the package allows one to fit multivariate normally distributed continuous and categorical longitudinal outcomes jointly with the time-to-event variable of interest \\ \hline
   Time-to-event submodel & Time-to-event submodel fitted with the \texttt{coxph($\cdot$)} function from the \textbf{survival} package. A number of relative risk and accelerated failure time survival model options are available, including Weibull, piecewise proportional hazards, Cox proportional hazards, and proportional hazards with a spline-approximated baseline risk function. The time-to-event submodel supported the stratification procedure\\ \hline
   Models comparison & Joint models comparison can be performed with the Deviance Information Criterion (DIC) available with the \texttt{anova($\cdot$)} function \\ \hline
   Post-fit analysis & Various post-fit tools for model diagnostic checks are available as diagnostic plots. Functionalities are available for computing dynamic predictions for both the longitudinal and time-to-event outcomes and assessment of model accuracy in terms of discrimination and calibration\\ \hline
      Extensions & Joint modeling for multivariate longitudinal outcomes and for time-varying association structures using P-splines are two very recent extensions implemented within the \textbf{JMbayes} package \\ \hline \\
\end{tabular}

\caption{Main functionalities of \texttt{R} package \textbf{JMbayes}.}
\label{Table JMbayes}
\end{table}

%---------------------------------------------------------------------

\begin{table}[!hb]
\centering
\setcellgapes{1pt}
\makegapedcells
\newcolumntype{L}[1]{>{\raggedright\arraybackslash }b{#1}}
\newcolumntype{C}[1]{>{\centering\arraybackslash }b{#1}}
%\begin{tabularx}{16cm}{L{4cm}|L{12cm}}
\begin{tabular}{L{3cm}|L{13cm}}

   \multicolumn{2}{c} {joineR \citep{joiner_2017,joineR2}} \\ \hline \hline

   Estimation method & Frequentist approach with maximum likelihood estimation using an EM algorithm \\ \hline
   Association structure &  The implemented associations structures are based on an extended version of the ``shared random effects'' model proposed by Wulfsohn and Tsiatis \citep{Wulfsohn1997} \\ \hline
   Longitudinal submodel & Longitudinal submodel is univariate and fitted with a linear mixed-effects model (splines and non normal responses are not feasible) \\ \hline
   Time-to-event submodel & Time-to-event submodel is a Cox proportional hazards model with log-Gaussian frailty\\ \hline
   Models comparison & Models comparison can be performed with Likelihood methods \\ \hline
   Post-fit analysis & Exact standard errors intervals can be obtained with implemented bootstrap methodology \\ \hline
      Extensions & The \textbf{joineRML} package \citep{hickey_joint_2016,hickey_joinerml:_2018} is a recent extension of the joineR which allows one to fit multivariate linear longitudinal data with a correlated time-to-event using Bayesian  Monte Carlo EM algorithm. \texttt{joineRML} also includes functions to derive dynamic predictions. \\ \hline \\
\end{tabular}
\caption{Main functionalities of \texttt{R} packages \textbf{joineR} and \textbf{joineRML}.}
\label{Table joineR}
\end{table}

%---------------------------------------------------------------------

\begin{table}[!hb]
\centering
\setcellgapes{1pt}
\makegapedcells
\newcolumntype{L}[1]{>{\raggedright\arraybackslash }b{#1}}
\newcolumntype{C}[1]{>{\centering\arraybackslash }b{#1}}
%\begin{tabularx}{16cm}{L{4cm}|L{12cm}}
\begin{tabular}{L{3cm}|L{13cm}}

   \multicolumn{2}{c} {lcmm \citep{proust2015lcmm}} \\ \hline \hline

   Estimation method & Frequentist approach based on maximum likelihood estimation using a modified Marquardt algorithm  \\ \hline
   Association structure &  The associations structures available are based on joint latent class models assumptions (joint models that consider homogeneous latent subgrouprocessing speed of individuals sharing the same longitudinal trajectory and risk of event)\\ \hline
   Longitudinal submodel & The linear model can include an univariate Gaussian outcome, an univariate curvilinear outcome, an univariate ordinal outcome and curvilinear multivariate outcomes\\ \hline
   Models comparison & Models comparison can be performed with Likelihood methods \\ \hline
   Post-fit analysis & Various post-fit functions with goodness-of-fit analyses, classification, plots, predicted trajectories, individual dynamic prediction of the event and predictive accuracy assessment are available \\ \hline \\

\end{tabular}
\caption{Main functionalities of \texttt{R} package \textbf{lcmm}.}
\label{Table lcmm}
\end{table}

%---------------------------------------------------------------------

\begin{table}[!hb]
\centering
\setcellgapes{1pt}
\makegapedcells
\newcolumntype{L}[1]{>{\raggedright\arraybackslash }b{#1}}
\newcolumntype{C}[1]{>{\centering\arraybackslash }b{#1}}
%\begin{tabularx}{16cm}{L{4cm}|L{12cm}}
\begin{tabular}{L{3cm}|L{13cm}}

   \multicolumn{2}{c} {\texttt{frailtypack}  \citep{rondeau2012frailtypack,Krol_2017}} \\ \hline \hline

   Estimation method & Frequentist maximum likelihood approach using either a parametric or semiparametric approach on the penalized likelihood for estimation of the hazard functions\\ \hline
   Longitudinal submodel and time-to-event submodel &  	Several type of joint models are implemented. In particular, joint models for recurrent events and a terminal event, for two time-to-event outcomes for clustered data, for two types of recurrent events and a terminal event, for a longitudinal biomarker and a terminal event and joint models for a longitudinal biomarker, recurrent events and a terminal event \\ \hline
   Models comparison & Two criteria for assessing model's predictive accuracy are implemented an can be used for models comparison\\ \hline
   Post-fit analysis & Each model function allows to evaluate goodness-of-fit analyses and provides plots of baseline hazard functions. Individual dynamic predictions of the terminal event and evaluation of predictive accuracy are also implemented \\ \hline
   \\
\end{tabular}
\caption{Main functionalities of \texttt{R} package \textbf{frailtypack}.}
\label{Table frailtypack}
\end{table}

%---------------------------------------------------------------------

\begin{table}[!hb]
\centering
\setcellgapes{1pt}
\makegapedcells
\newcolumntype{L}[1]{>{\raggedright\arraybackslash }b{#1}}
\newcolumntype{C}[1]{>{\centering\arraybackslash }b{#1}}
%\begin{tabularx}{16cm}{L{4cm}|L{12cm}}
\begin{tabular}{L{3cm}|L{13cm}}

   \multicolumn{2}{c} {\texttt{rstanarm} \citep{rstanarm}} \\ \hline \hline

   Estimation method & Bayesian approach using a MCMC algorithm with the function \texttt{stan\_jm($\cdot$)}. A multiple chain methodology is implemented (long chain is also feasible). Very clear and user friendly implementation\\ \hline
   Association structure &  Current value, current value and slope, shared random effects, lagged effects, cumulative effects and interaction effect associations structures are implemented\\ \hline
   Longitudinal submodel & Generalized linear mixed-effects model (the response has to belong to an exponential family distribution). The longitudinal part can be multivariate, contonuous and/or categorical. Linear slope, cubic splines and polynomial terms are allowed \\ \hline
   Time-to-event submodel &  The baseline hazard can be specified parametrically or non parametrically. A stratification procedure is not allowed\\ \hline
   Models comparison & There is currently no models comparison procedure implemented  \\ \hline
   Post-fit analysis & Several post-fit functions for dynamic predictions of the terminal event and longitudinal trajectories and as well as visualisation functions are available \\ \hline\\

\end{tabular}
\caption{Main functionalities of \texttt{R} package \textbf{rstanarm}.}
\label{Table rstanarm}
\end{table}

%---------------------------------------------------------------------

\begin{table}[!hb]
\centering
\setcellgapes{1pt}
\makegapedcells
\newcolumntype{L}[1]{>{\raggedright\arraybackslash }b{#1}}
\newcolumntype{C}[1]{>{\centering\arraybackslash }b{#1}}
%\begin{tabularx}{16cm}{L{4cm}|L{12cm}}
\begin{tabular}{L{3cm}|L{13cm}}

   \multicolumn{2}{c} {\texttt{bamlss} \citep{umlauf_bamlss_2018}} \\ \hline \hline

   Estimation method & Bayesian method \\ \hline
   Association structure &   Flexible additive joint models are implemented \\ \hline
   Longitudinal submodel & Univariate continuous longitudinal outcome is allowed an\\ \hline
   Time-to-event submodel &  A single time-to-event outcome is allowed \\ \hline\\

\end{tabular}
\caption{Main functionalities of \texttt{R} package \textbf{bamlss}.}
\label{Table bamlss}
\end{table}

\end{document}